\documentclass[aps,prl,twocolumn,floatfix,superscriptaddress,nofootinbib]{revtex4}

\usepackage{amssymb}
\usepackage{graphicx,subfigure}
\usepackage{bm}
\usepackage{amsfonts}
\usepackage{amsmath}
\usepackage{tabularx}

\def\C60{A$_x$C$_{60}$}

\def\HgCu3{HgCa$_2$Cu$_3$O$_{8+y}$}
\def\HgCu4{HgBa$_2$Ca$_3$Cu$_4$O$_{10+y}$}
\def\TlCu{Tl$_2$Ba$_2$CuO$_{6+\delta}$}
\def\TlCu3{Tl$_2$Ba$_2$Ca$_2$Cu$_3$O$_{10+y}$}
\def\TlCu4{Tl$_2$Ba$_2$Ca$_3$Cu$_4$O$_{12+y}$}

\def\BiCu3{Bi$_2$Sr$_2$Ca$_{2}$Cu$_3$O$_y$}

\def\8LSCO{La$_{1.88}$Sr$_{.12}$CuO$_4$}
\def\110LNSCO{La$_{1.5}$Nd$_{0.4}$Sr$_{0.1}$CuO$_{4}$}
\def\stage4LCO{La$_{2}$CuO$_{4+\delta}$}
\def\Y248{YBa$_2$Cu$_4$O$_8$}

\def\NbSe2{NbSe$_2$}
\def\TaSe2{TaSe$_2$}
\def\TiSe2{TiSe$_2$}
\def\NaCoOH2O{Na$_{0.3}$CoO$_{2y}$H$_2$O}
\def\MgB2{MgB${}_2$}

\def\URu2Si2{URu$_2$Si$_2$}

\begin{document}

\title{Pair Density Wave correlations in the Kondo-Heisenberg Model}
\author{ Erez Berg}
\affiliation{Department of Physics, Harvard University, Cambridge, Massachusetts 02138,
USA}
\author{ Eduardo Fradkin}
\affiliation{Department of Physics, University of Illinois at Urbana-Champaign, Urbana,
Illinois 61801-3080, USA}
\author{ Steven A. Kivelson}
\affiliation{Department of Physics, Stanford University, Stanford, California 94305-4060,
USA}
\date{\today }

\begin{abstract}
We show, using density matrix renormalization group calculations
complemented by field theoretic arguments, that the spin gapped
phase of the one dimensional Kondo-Heisenberg model exhibits
quasi-long range superconducting correlations
{\it only} at a non-zero
momentum. The local correlations in this phase resemble those of
the pair density wave state which was recently proposed to
describe the phenomenology of the striped ordered high temperature superconductor {La$_{2-x}$Ba$_x$%
CuO$_4$}, in which the spin, charge, and superconducting orders
are strongly intertwined.
\end{abstract}

\maketitle

Recent experiments in the {high temperature superconductor} {La$_{2-x}$Ba$_x$%
CuO$_4$} near doping $x=1/8$ have revealed a dramatic layer
decoupling
effect 
in which anomalous mesoscopic 2D superconductivity 
persists well above the macroscopic 3D superconducting transition
temperature, $T_c$.
\cite{li-short,tranquada-short,schafgans-short} Moreover, the
superconductivity coexists 
with static stripe (charge and spin) order.
It has been proposed that the anomalous superconducting properties
are evidence of
the existence of a novel type of superconducting state, the
pair-density
wave (PDW). 
\cite{berg-short,berg-2008a, berg-short-b}

The PDW is a state in which charge, spin and superconducting (SC)
orders are
\textit{intertwined} in a spatially modulated fashion. 
The SC order has a wave vector $\mathbf{Q}$ which is the same as that of the
spin density wave (SDW) 
and half of the ordering wave vector $2\mathbf{Q}$ of the charge density
wave (CDW). 
Its SC order 
is 
Larkin-Ovchinnikov-like, 
but without the magnetization of the latter. 
Although much is known about the 
properties of this state\cite%
{berg-2008a,berg-short-b}, 
there is, as yet, no 
fully satisfactory microscopic theory. 

In the context of Bardeen-Cooper-Schrieffer type mean-field 
theories, 
a PDW is only ever stable at strong coupling\cite{loder-2009} (
{\it i.e.}
outside the regime in which such 
treatments
are reliable).
Slave-boson mean-field theories of the $t-J$ model 
find that, although the PDW is quite competitive
energetically, it (barely) loses to the uniform d-wave SC state\cite%
{yang-short}.
While early numerical variational Monte Carlo studies 
of the $t-J$ model
found a regime in which the PDW appeared to be 
stable\cite{himeda-2002}, more recent studies have 
found that it
has 
slightly higher variational energy than 
the uniform d-wave state\cite{raczkowski-short,capello-2008}.

In the present paper we study the superconducting correlations in the 1D
Kondo-Heisenberg model (KHM). This is the simplest model in which one can
investigate the interplay between strong antiferromagnetic ordering
tendencies, represented by a Heisenberg chain, and possible superconducting
and charge-density wave orders, derived from an itinerant electron band to
which it is coupled. 
The 1D character of the model permits us to employ the powerful
numerical density-matrix renormalization group
(DMRG)\cite{white-1992a} and analytic bosonization methods to
solve the problem, despite the strong interactions. On the down
side, there are special features of 1D physics, which may raise
questions concerning the applicability of
the results to higher dimensional
situations. On the other hand, especially since the order we are
investigating is unidirectional, and thus has an essentially 1D geometry, it
is plausible that the local structure of the correlations up to intermediate
scales are dimension independent.

The key finding from our DMRG studies is that, for the range of
parameters considered here, the 1D KHM exhibits a spin-gapped
phase with quasi-long range (power-law) PDW correlations,{\it
i.e.} superconducting correlations which oscillate with a period
$2b$ where $b$ is the lattice constant of the Heisenberg chain. At
the same time the uniform singlet superconducting correlations are
small and apparently fall exponentially with distance. Since the
same model exhibits substantial, although short-ranged correlated,
antiferromagnetic tendencies with the same period, this state can
clearly be identified as a fluctuating version of the long-sought
PDW. Note that the occurrence of a spin-gap in the 1D
Kondo-Heisenberg model has been discussed
insightfully in the literature\cite{white-1996,sikkema-1997}
and the possibility of an oscillatory superconducting order
parameter was previously inferred on the basis of bosonization
studies.\cite{zachar-1996,zachar-1999,coleman-short,moreno-short,
zachar-2001,zachar-2001b} However, we believe that this is the
first place in which the existence and character of this state has
been derived from a microscopic model, and the nature of the
correlations is elucidated.\cite{note-batista}

\paragraph{Model:}
The 1D KHM\ is defined as a one dimensional electron gas (1DEG) coupled to a
spin-$\frac{1}{2}$ chain:
\begin{equation}
H=H_{\mathrm{1DEG}}+H_{\mathrm{Heis}}+H_{\mathrm{K}}  \label{H}
\end{equation}%
where
\begin{eqnarray}
H_{\mathrm{1DEG}} &=&-t\sum_{j,\sigma }c_{j\sigma }^{\dagger }c_{j+1\sigma }+%
\text{\textrm{h.c.}}-\mu \sum_{j,\sigma }n_{j}\text{,}  \label{1deg} \\
H_{\mathrm{Heis}} &=&J_{H}\sum_{j}\mathbf{S}_{j}\cdot \mathbf{S}_{j+1}\text{,%
}  \label{heis} \\
H_{\mathrm{K}} &=&J_{K}\sum_{j,a}S_{j}^{a}\left[ c_{j\sigma }^{\dagger
}(s^{a})_{\sigma \sigma ^{\prime }}c_{j\sigma ^{\prime }}\right] \text{.}
\label{kondo}
\end{eqnarray}%
Here, $c_{j\sigma }^{\dagger }$ creates an electron with spin $\sigma $ at
site $j$, $\mathbf{S}_{j}$ is the spin $\frac{1}{2}$ operator of the spin
chain, and $s^{a}=\frac{1}{2}\tau ^{a}$ ($\tau ^{a=x,y,z}$ are Pauli
matrices).

In typical physical circumstances in which Kondo physics arises,
one would expect $J_H$ and $J_K \ll t$.  In this limit, the length
scales  characterizing the Kondo effect are exponentially 
large, and hence not readily accessible by any numerical method.
We
therefore 
use $J_H \sim J_K \sim t$.  
On the basis of the field-theoretic analysis (see below), we
expect the character of the
phases 
to survives to small $J/t$.
Moreover, the $J_H \sim J_K \sim t$ regime is not necessarily
unphysical; it can be derived from the $U\to\infty$ limit of a
Hubbard model on the spin-chain, with chemical potential chosen so
that there is one electron per site, and with hopping matrix
element along the chain, $t_H = \sqrt{J_H U/2}$, and hopping
between the spin-chain and the 1DEG, $t_K=\sqrt{J_KU/2}$.

\begin{figure}[t]
\includegraphics[width=0.25\textwidth]{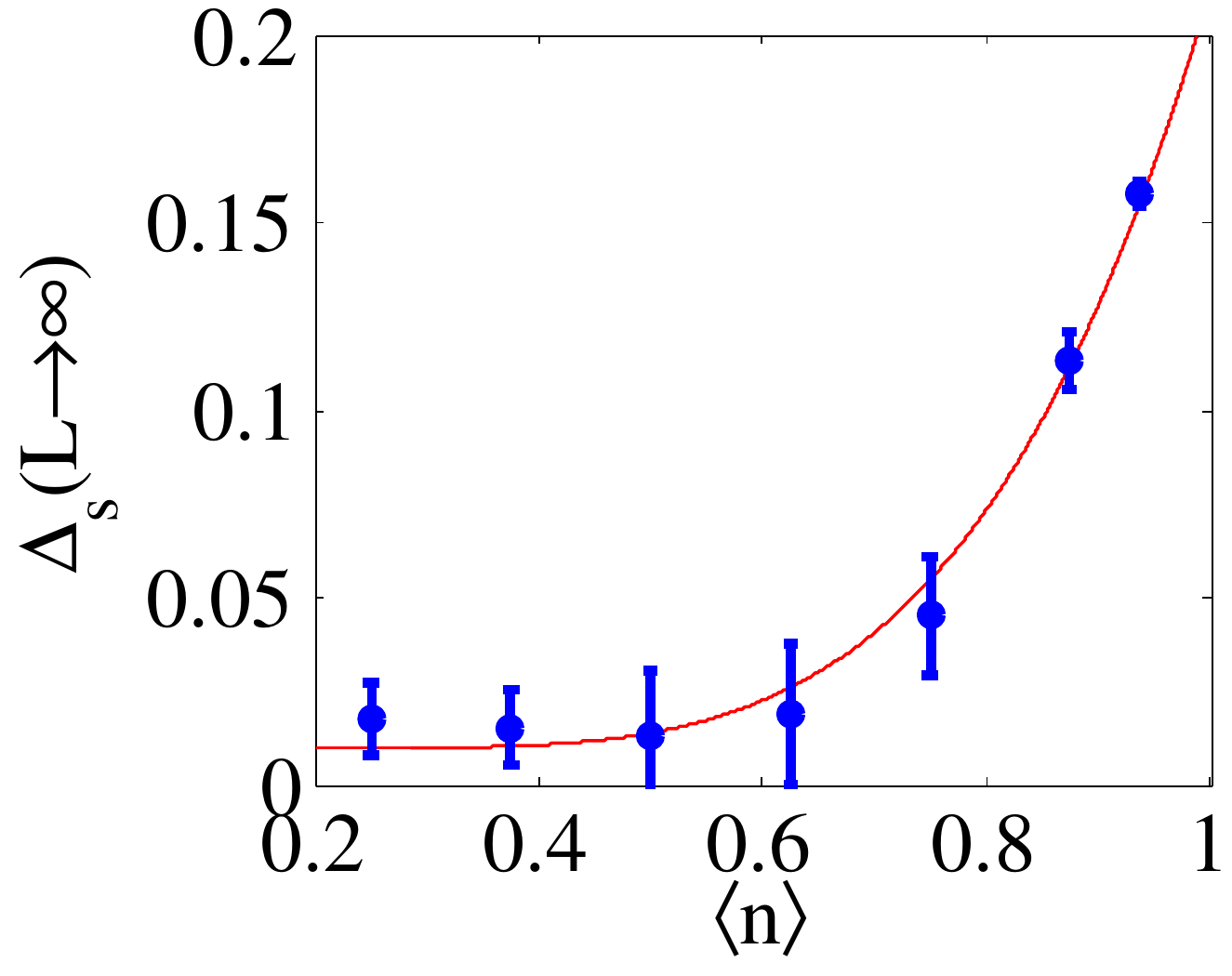}
\vspace{-0.1in} \caption{(Color online.) The spin gap vs. the
electron concentration in the 1DEG. $J_H=J_K=2t$. The error bars
are a result of the extrapolation to the thermodynamic limit.
(Relative to the extrapolation error, the DMRG truncation error is
negligible.)} \label{fig:sg}
\end{figure}

\paragraph{Numerical results:}
The model (\ref{H}) was solved using DMRG on finite lattices with
$L=32-128$ and open boundary conditions. Up to $m=1800$ states
were kept, giving DMRG truncation errors smaller than $10^{-6}$.

Fig. \ref{fig:sg} shows the spin gap $\Delta_{s}=E_{0}\left(1\right) -E_{0}\left(0\right) $, where $%
E_{0}\left( S_{z}\right) $ is the ground state energy of a system
with a $z$ spin projection $S_{z}$. The spin gap was extrapolated
to the thermodynamic ($L\rightarrow \infty $) limit. The results are shown for $%
J_{H}=J_{K}=2t$, as a function of the concentration of electrons
in the 1DEG, $n$. Due to the particle-hole symmetry of the model,
it is sufficient to consider $n<1$.

Near $n=1$ there is a sizable spin gap\cite{sikkema-1997};
$\Delta_s$ decreases away from $n=1$. 
For $n=1$ (not shown), 
the spin gap is $\Delta _{s}\approx 0.8$,
but there is also a finite charge gap
in the $L\rightarrow\infty$ limit. 
We henceforth focus on 
$n=0.875$, for which $\Delta_s$ is substantial. Since $\Delta_s$
persists at lower densities, we expect the low-energy properties
at smaller $n$ to be similar, although the
correlation length is 
larger.

\begin{figure}[t]
\includegraphics[width=0.52\textwidth]{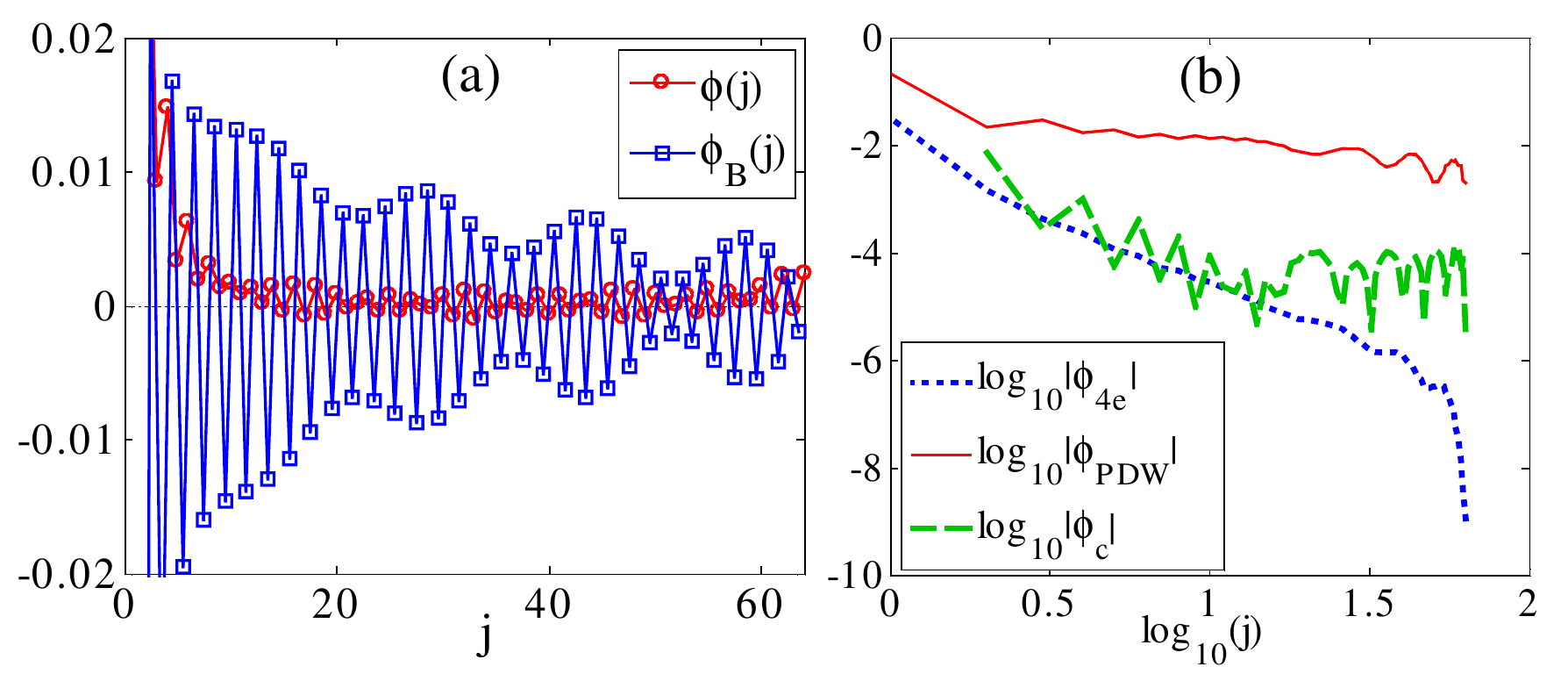}
\vspace{-0.3in} \caption{(Color online.)(a) The SC order
parameters $\phi$ and $\phi_B$ (see text) as a function of
position in an $L=64$ system with $n=0.875$. (b) Measurements of
$\phi_{\mathrm{PDW}}$, $\phi_{4e}$, $\phi_{c}$ vs. position. The
oscillatory behavior close to the right boundary is an edge
effect.} \label{fig:op}
\end{figure}

\paragraph{PDW correlations:}
The opening of a spin gap is expected to lead to enhanced SC (as
well as CDW) correlations. To study these correlations, we have
applied a local pair field  to the left
boundary\cite{feiguin-2008-short}:
\begin{equation}
H_{\text{\textrm{pair}}}=\Delta (c_{1\uparrow }^{\dagger }c_{2\downarrow
}^{\dagger }-c_{1\downarrow }^{\dagger }c_{2\uparrow }^{\dagger })+\text{%
\textrm{h.c.}}  \label{pf}
\end{equation}%
where we fixed $\Delta =0.5t$. (We have checked explicitly that
the results do not depend on the size of $\Delta$.) The
superconducting response of the system was probed by measuring the
following induced order parameters throughout the
system\cite{comment-exponent}:
\begin{equation}
\phi \left( j\right)  =\langle c_{j\uparrow }^{\dagger
}c_{j\downarrow }^{\dagger }\rangle \text{,} \quad \phi _{B}\left(
j\right)  =\frac{1}{2}\langle c_{j\uparrow }^{\dagger
}c_{j+1\downarrow }^{\dagger }-c_{j\downarrow }^{\dagger
}c_{j+1\uparrow }^{\dagger }\rangle \text{,}
\end{equation}%
where $\phi(j)$ and $\phi_B(j)$ are, respectively, the expectation
of the singlet pair creation operator on site $j$ and on the bond
from site $j$ to site $j+1$. Fig. \ref{fig:op}a shows $\phi(j)$
and $\phi_B(j)$ in an $L=64$ system. $\phi (j)$ appears to decay
very rapidly away from the left boundary. $\phi_{B}(j)$ decays
much more slowly, and exhibits pronounced oscillations as a
function of position with wavevector 
$q=\pi/b $, as it changes its sign between every consecutive
bonds. Longer periods are also apparent in the figure. These
oscillations clearly indicate that the dominant pairing
correlations are at a non-zero momentum.

Refs. \cite{zachar-1996,zachar-1999,coleman-short,
zachar-2001,zachar-2001b} proposed, based on bosonization, that
the spin-gapped phase of the KHM has dominant pairing correlations
at a non-zero wavevector, described by a \textquotedblleft
composite\textquotedblright\ order parameter\cite{zachar-2001b}
\begin{equation}
\phi _{c}\left( j\right) =\left( -1\right) ^{j}\langle \lbrack \sum_{\sigma
,\sigma ^{\prime }}c_{j-1\sigma }^{\dagger }\left( is^{y}\mathbf{s}\right)
_{\sigma \sigma ^{\prime }}c_{j+1\sigma ^{\prime }}^{\dagger }]\cdot \mathbf{%
S}_{j}\rangle \text{.}
\end{equation}%
In addition, PDW order should be accompanied by a uniform ($q=0$) \textquotedblleft charge 4%
$e$\textquotedblright\ order parameter\cite{berg-2009}:
\begin{equation}
\phi _{4e}\left( j\right) =\langle c_{j\uparrow }^{\dagger }c_{j\downarrow
}^{\dagger }c_{j+1\uparrow }^{\dagger }c_{j+1\downarrow }^{\dagger }\rangle
\text{.}
\end{equation}%
Fig \ref{fig:op}b shows $\phi _{\mathrm{PDW}}\left( j\right)
\equiv \left( -1\right)
^{j}\phi _{B}\left( j\right) $, as well as $\phi _{c}\left( j\right) $ and $%
\phi _{4e}\left( j\right) $, as a function of position, on a
logarithmic scale. The largest, and most slowly decaying, order
parameter is $\phi _{\mathrm{PDW}}\left( j\right) $, suggesting
that the
system is best described as a fluctuating 
PDW state.
As expected, $\phi _{4e}\left( j\right) $ and $\phi _{c}\left(
j\right) $ are non-zero, but 
small. $\phi _{c}\left( j\right)
$ is modulated as a function of position, while $\phi _{4e}\left(
j\right) $ is smooth.
\begin{figure}[t]
\includegraphics[width=0.45\textwidth]{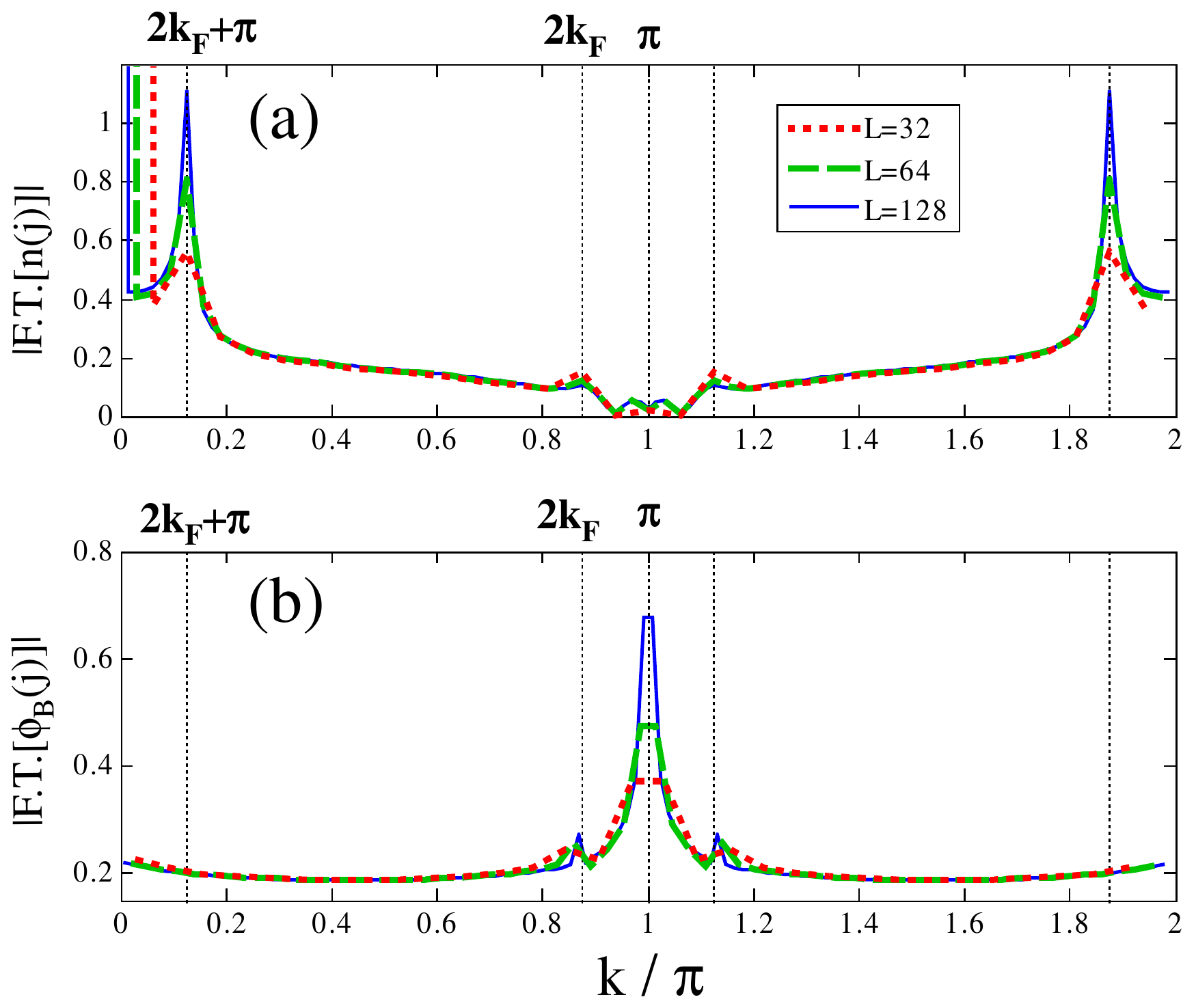}
\vspace{-0.15in} \caption{(Color online.) Absolute values of the
Fourier transforms of (a) the charge [$\langle n(j) \rangle$] and
(b) the SC [$\phi_{B}(j)$] orders. $L=32$--$128$.}
\label{fig:fourier}
\end{figure}

The wavevectors of the leading SC and CDW fluctuations can be
determined by a Fourier analysis of the SC\ and CDW\
orders.\cite{comment-white-2002} Fig. \ref{fig:fourier} shows the
absolute values of the Fourier transforms of $\phi _{B}\left(
j\right) $ and $n\left( j\right) \equiv\sum_{\sigma }c_{j\sigma
}^{\dagger }c_{j\sigma }$ for system sizes between $L=32$ and
$128$. The charge density exhibits
a large peak at 
$q=2k_{F}+\pi / b$
which grows as a function of system size, where $2k_{F}\equiv\pi
n$. There are also small sub-leading features at $q=2k_{F}$ and
$q=\pi/b$. The main feature in the Fourier transform of $\phi _{B}$
is a pronounced peak at 
$q=\pi/b$, with a sub--leading peak at
$q=2k_{F}$. This shows unambiguously that the dominant order in
this system is a PDW\ with 
$q=\pi/b$, accompanied by CDW\
correlations at 
 $q=2k_{F}+\pi /b$.

\begin{figure}[t]
\includegraphics[width=0.54\textwidth]{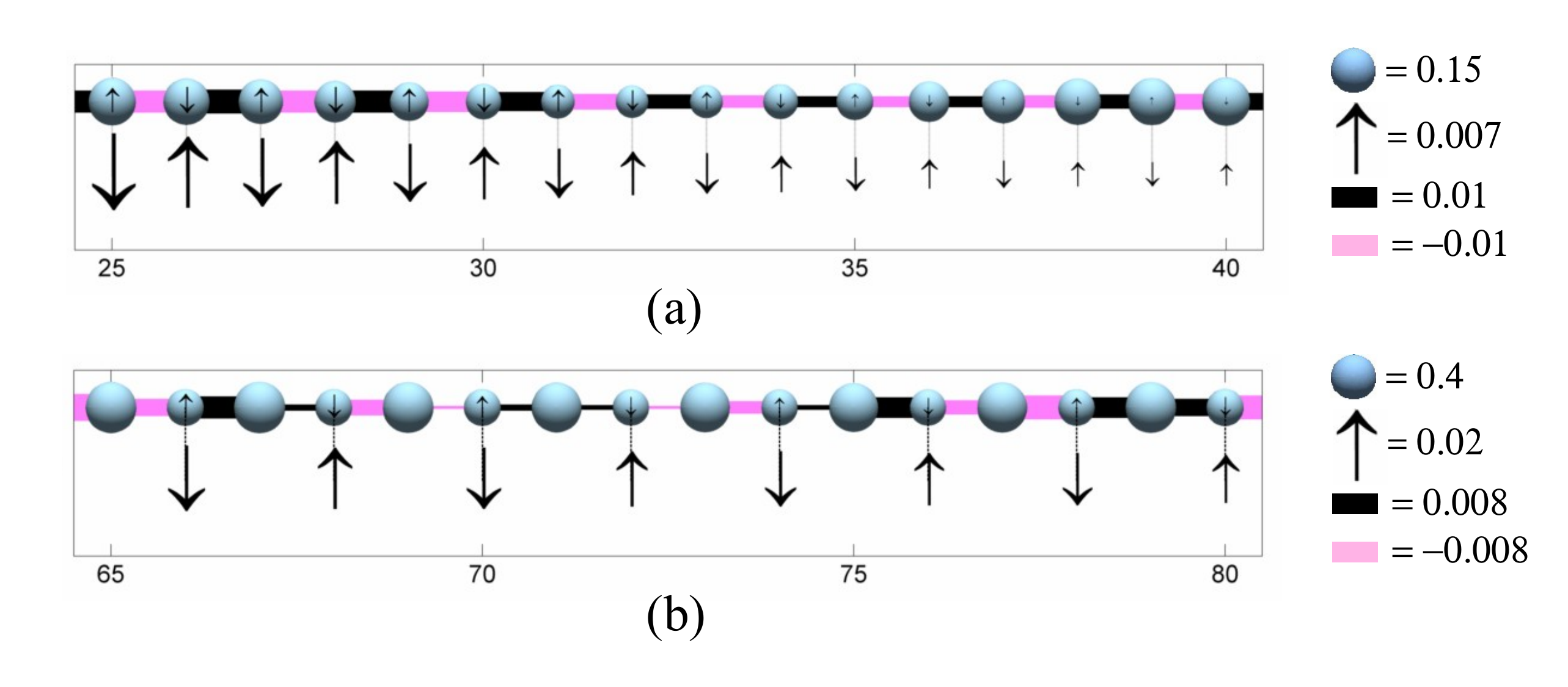}
\vspace{-0.4in} \caption{
(Color online.) Order parameters in (a) an $L=64$ KHM chain with
$n=0.875$, and (b) an $L=128$ ``diluted'' KHM chain (with one spin
site for each two 1DEG sites) with $n=0.625$. Circles: 1DEG hole
density ($1-\langle n(j) \rangle$), bond color/thickness: the
bond-centered SC amplitude $\phi_B(j)$. The arrows show the spin
density $\langle S^z(j) \rangle$.} \label{fig:corr}
\end{figure}

In order to elucidate further the nature of the microscopic
correlations in the system, we perform another simulation in which
both a pair field [Eq (\ref{pf})] and a Zeeman field,
$H_{\mathrm{Z}}=-hS_{j=1}^{z}$ ($h=0.5t$) are applied to the left
boundary of the system. The induced charge, superconducting, and
magnetic ($\langle S^z_j \rangle$) order parameters are shown near
the middle of the $L=64$ system in Fig. \ref{fig:corr}a. The
magnetic order oscillates at wavevector  $q=\pi /b$ (the same as
the PDW) with an envelope that decays
exponentially on longer length scales.

Next, we would like to understand what determines the PDW\
wavevector. We performed another calculation in which the spin
chain is \textquotedblleft diluted\textquotedblright , {\it i.e.} there
is one spin site for every \emph{two} 1DEG
sites. {}Eq. (\ref{heis}) is replaced by $\tilde{H}_{\mathrm{Heis}%
}=J_{H}\sum_{j}\mathbf{S}_{2j}\cdot \mathbf{S}_{2j+2}$, and similarly $%
\tilde{H}_{\mathrm{K}}=J_{K}\sum_{j,a}S_{2j}^{a}[c_{2j\sigma }^{\dagger
}(s^{a})_{\sigma \sigma ^{\prime }}c_{2j\sigma ^{\prime }}]$. Fig. \ref%
{fig:corr}b illustrates the results for $\langle S_j^z\rangle$,
$\langle n_j \rangle$, and $\phi_B$ near the middle of an $L=128$
system. (In order to maintain a large spin gap, $n$ was 
taken to be $0.625$.) Clearly, the PDW order changes sign
across every spin
site, indicating that the dominant PDW wavevector is
again $q=\pi/b$,
where now $b=2$.
Thus, the period of the PDW is tied to that of the local
(fluctuating) magnetic
ordering.
The local correlations in Fig. \ref{fig:corr}a-b are a one
dimensional version of the
phenomenologically proposed ``striped-superconducting'' state for
La$_{2-x}$Ba$_x$CuO$_4$ \cite{berg-short}.


\paragraph{Continuum limit:}

Analytical progress can be made in the limit $J_{H}$,$t\gg J_{K}$,
where we may first take the continuum limit of both the 1DEG and
the spin chain. We use a description in terms of the Bosonic
fields
${\varphi _{c},\varphi _{s}}$ and ${\tilde{\varphi}_{s}}$,
representing charge/spin fluctuations in the 1DEG and spin chain,
respectively [and the respective conjugate fields $\theta _{c}$,
$\theta _{s}$ 
and $\tilde{ \theta}_{s}$]. The Hamiltonian densities of the 1DEG
and the spin chain take the form
\cite{white-1996,zachar-2001,zachar-2001b}
\begin{eqnarray}
&&\mathcal{H}_{\mathrm{1DEG}}=\sum_{\alpha =c,s}\;v_{\alpha
}\;[\frac{K_{\alpha }}{2}\left( \partial _{x}\theta _{\alpha
}\right) ^{2}+\frac{1}{2K_{\alpha }}\left( \partial _{x}\varphi
_{\alpha }\right) ^{2}]  \nonumber \\
&&\mathcal{H}_{\mathrm{Heis}}=\frac{1}{2}\;\tilde{v}_{s}\;[(\partial _{x}\tilde{%
\theta}_{s})^{2}+(\partial _{x}\tilde{\varphi}_{s})^{2}]
\label{Heis}
\end{eqnarray}
where $K_{c}$, $K_{s}$, $v_{c}$, $v_{s}$ and $\tilde{v}_{s}$ are,
respectively, the charge and spin Luttinger parameters of the
1DEG, and the corresponding charge and spin velocities. The
various bosonized fields satisfy the commutation relation $\left[
\varphi _{\alpha }\left( x\right) ,\partial _{x}\theta _{\alpha
}\left( x^{\prime }\right) \right] =i\delta
\left( x-x^{\prime }\right) $, and similarly for $\tilde{\theta}_{s}$, $%
\tilde{\varphi}_{s}$. We neglect marginally irrelevant
contributions to $H_{\mathrm{Heis}}$.

For a an incommensurate filling $n$ of the 1DEG, only ``forward
scattering'' terms in the spin channel can couple the 1DEG and the
spin chain. 
Up to irrelevant (backscattering) operators, the Kondo Hamiltonian
density is
$\mathcal{H}_{\mathrm{K}} = \frac{J_{K}a}{8\pi }[\left(
\partial _{x}\varphi
_{+}\right) ^{2}-\left( \partial _{x}\varphi _{-}\right)
^{2}]+\mathcal{H}_{\mathrm{int}}$
\cite{white-1996}
where 
$\theta _{\pm }=\frac{1}{\sqrt{2}}( \tilde{%
\theta}_{s}\pm \theta _{s}) $, $\varphi _{\pm }=\frac{1}{\sqrt{2}}%
( \tilde{\varphi}_{s}\pm \varphi _{s})$, and
$\mathcal{H}_{\mathrm{int}}=\frac{\cos ( \sqrt{4\pi }\theta
_{-})}{2\left( \pi a\right) ^{2}} \lbrack \cos (\sqrt{4\pi
}\varphi _{-})+\cos (\sqrt{4\pi }\varphi _{+})]$. ($a$ is a
microscopic
cutoff.) Under 
renormalization, 
$\cos ( \sqrt{4\pi }\theta _{-}) \cos ( \sqrt{4\pi }\varphi _{+})
$ is marginally relevant\cite{white-1996}, while $\cos (
\sqrt{4\pi }\theta _{-}) \cos ( \sqrt{4\pi }\varphi _{-}) $ is
irrelevant, since it contains the dual fields $\theta _{-}$ and
$\varphi _{-}$. The strong coupling phase has a spin gap, while
the charge degree of freedom $\varphi _{c}$ remains decoupled and
gapless.

\paragraph{Correlations in the spin gapped phase:}
The form of the dominant (slowest decaying) correlations follows
from the following considerations. A theorem by Yamanaka \emph{et.
al.} \cite{yamanaka-1997}
guarantees the existence of a charge zero, momentum $2k_{F}^{\ast }=\pi n_{%
\text{\textrm{tot}}}$ gapless excitation, where $n_{\mathrm{tot}}$
is the total electron density in the system (counting both the
1DEG and the spin chain). Here,
$n_{\text{\textrm{tot}}}=n+\frac{1}{b}$, therefore $2k_{F}^{\ast
}=2k_{F}+\pi /b$.
Let us denote the operator that creates
these excitations $\hat{O}_{2k_{F}^{\ast }}$. Since there is a
spin gap, $\hat{O}_{2k_{F}^{\ast }}$ is necessarily a spin
singlet, \emph{i.e.}, 
a CDW operator.

In addition, as long as there is no charge gap, the singlet
\textquotedblleft $\eta -$pairing operator\textquotedblright \cite
{zachar-1999,zachar-2001b}
$\hat{O}_{\eta }=\psi _{-\uparrow }\psi _{-\downarrow }=\frac{1}{2\pi a}\exp[i%
\sqrt{2\pi }\left( \theta _{c}-\phi _{c}\right) ]$,
also creates gapless excitations. ($\psi _{\pm ,\sigma }$
annihilate right/left moving electrons with spin $\sigma =\uparrow
,\downarrow $, respectively.) This operator has total momentum
$-2k_{F}$ and charge $2e$. Therefore, the \textquotedblleft PDW
operator\textquotedblright\ $\hat{O}_{\text{\textrm{PDW
}}}=\hat{O}_{\eta }\hat{O}_{2k_{F}^{\ast }}$ also creates gapless
excitations. Adding the quantum numbers carried by $\hat{O}_{\eta }$ and $%
\hat{O}_{2k_{F}^{\ast }}$, we see that
$\hat{O}_{\text{\textrm{PDW}}}$ carries charge $2e$ and momentum
$\pi /b$. This guarantees the existence of quasi-long range PDW\
correlations in the spin gapped phase. As usual, the correlations
of $\hat{O}_{\mathrm{PDW}}$ (as well as those  of
$\hat{O}_{2k_F^*}$) fall off with a non-universal exponent, which
depends on $K_c$.
The (zero momentum) Cooper pair
operator  is $\hat{O}_{SC}=\psi _{+\uparrow }\psi _{-\downarrow }=\frac{1}{2\pi a}e^{i%
\sqrt{2\pi }\left( \theta _{c}+\phi _{s}\right) }$.
Its correlations are short ranged, since $\phi
_{s}=\left( \phi _{+}+\phi _{-}\right) /\sqrt{2}$;
in the spin
gapped phase the field $\theta _{-}$ is pinned,
while its dual $\phi_{-}$ undergoes strong fluctuations,
suppressing the correlations of $\hat{O}_{SC}$.
Consequently, the leading
superconducting correlations
are for operators with non-zero
momentum.

Generically, any singlet operator that carries charge
$2e$ and
momentum $\pi /b$ is expected to couple to
$\hat{O}_{\text{\textrm{PDW}}}$, and therefore
to have quasi-long range correlations. For example, both $\phi _{\text{%
\textrm{PDW}}}$ and $\phi _{c}$ defined above have the correct
quantum numbers, and therefore their correlations should fall off
with the same exponent as that of $\hat{O}_{\textrm{PDW}}$.
According to our numerical simulations, the spin gapped phase has
strong PDW correlations, so it is best characterized by the $\phi _{\text{%
\textrm{PDW}}}$ order parameter.
The results in Fig. \ref{fig:fourier} are fully
consistent with the field theoretic analysis above. In particular,
the density profile shows a large peak at $q=2k_{F}^{\ast }$ which
grows with system size, indicating slowly decaying fluctuations
centered at that wavevector. The pairing correlations are strongly
peaked at 
$q=\pi/b $, with a subdominant peak
 (which does not grow with $L$)
at $q=2k_{F}$,
corresponding to the gapless $\eta $ pairing mode.

\paragraph{Discussion:}
The correlations in the spin gapped phase of the
1D KHM
are best described as a PDW phase, which is a (quasi-)condensate
of non-zero center of mass momentum Cooper pairs. Locally, the
correlations are strikingly similar to those of the PDW state
recently proposed to describe the striped phase of
La$_{2-x}$Ba$_{x}$CuO$_{4}$, which intertwines spin, charge, and
density orders. A study of a two-chain KHM 
found,
instead, dominant uniform 
pairing
correlations\cite{xavier-2008}.
It remains an important question whether the PDW state survives in other multi-chain generalizations of the present model. Finally, the 1D KHM\ can be
viewed as a variation of the three-band copper-oxide
model\cite{emery-1987}, with strongly localized spins on the Cu
sites and a 1DEG\ representing doped holes on O sites. Therefore
it seems plausible that such a model can exhibit a PDW\ phase as
well. Whether it can be realized in the physically relevant
parameter regime remains to be seen.


\begin{acknowledgments}
We thank P. Coleman, T. Giamarchi, A. Tsvelik and S. White for
discussions. We
thank the 
KITP at UCSB  for hospitality.
This work was supported in part by the NSF, under grants DMR-
0758462 (EF), DMR-0531196 (SAK), DMR-0705472 and DMR-0757145 (EB),
and PHY05-51164 at KITP (EB, EF, SAK), and by the
DOE under Contracts DE-FG02-07ER46453 
at UIUC (EF) and DE-FG02-06ER46287 
at Stanford (SAK).
\end{acknowledgments}

\bibliography{pdw-kh}

\begin{widetext}
\appendix
\section{Supplementary information}
Throughout the paper, we have used various boundary perturbations
(either pairing fields, Zeeman fields, or the boundary itself) to
excite various sorts of order parameters. The decay of these order
parameters into the bulk gives information on the leading response
of the system. In order to verify that our results (in particular,
the fact that the dominant pairing fluctuations are at a non--zero
wavevector) are not induced by the external boundary
perturbations, we have made the following additional checks:

\begin{enumerate}
\item We have varied the strength of the bond--centered boundary
pair field, $H_{\mathrm{pair}}$, in the range $\Delta =0.1-0.5t$;

\item We tried to apply an on--site pair field on the first site,
rather than a bond--centered pair field, in order to verify that
the nature of the correlations is not sensitive to the form of the
pair fields; and

\item We set $\Delta =0$ and calculated the equal--time pair--pair
correlation function.
\end{enumerate}

In the this Supplementary Information, we describe these results
of these calculations. In all cases, the doping of the 1D electron
gas was $n=1.125$ particles per site, and the following parameters
were used: $J_{H}/t=J_{K}/t=2$.

\subsection{Dependence on the boundary pair field strength}

Fig. \ref{fig1} shows the induced pair fields\qquad
\begin{eqnarray}
\phi \left( j\right) &=&\langle c_{j\uparrow }^{\dagger
}c_{j\downarrow
}^{\dagger }\rangle  \notag \\
\phi _{B}\left( j\right) &=&\frac{1}{2}\langle c_{j\uparrow
}^{\dagger }c_{j+1\downarrow }^{\dagger }-c_{j\downarrow
}^{\dagger }c_{j+1\uparrow }^{\dagger }\rangle \text{,}\label{OP}
\end{eqnarray}%
in three calcalutions, in which a bond--centered pair field
\begin{equation} H_{\mathrm{pair}}=\Delta (c_{j\uparrow
}^{\dagger }c_{j+1\downarrow }^{\dagger }-c_{j\downarrow
}^{\dagger }c_{j+1\uparrow }^{\dagger } + H.c.)\end{equation} was
applied to the first bond
with a varying strength: $\Delta /t=0.1,0.2,0.5$. The system sizes were $%
L=64 $ in all calculations. The results are qualitatively similar
to each other. In particular, the fact that the dominant response
is at a non--zero wavevector $q=\pi /a$ does not depend on the
value of $\Delta /t$. Note that, between $\Delta /t=0.1$ and
$0.2$, the response is roughly linear in the applied field.
Between $\Delta /t=0.2$ and $0.5$, the response nearly saturates.
In all cases, we found that the on--site order parameter $\phi
\left( j\right) $ (not shown) decays very rapidly.
\begin{figure}[h]
\includegraphics[width=0.35\textwidth]{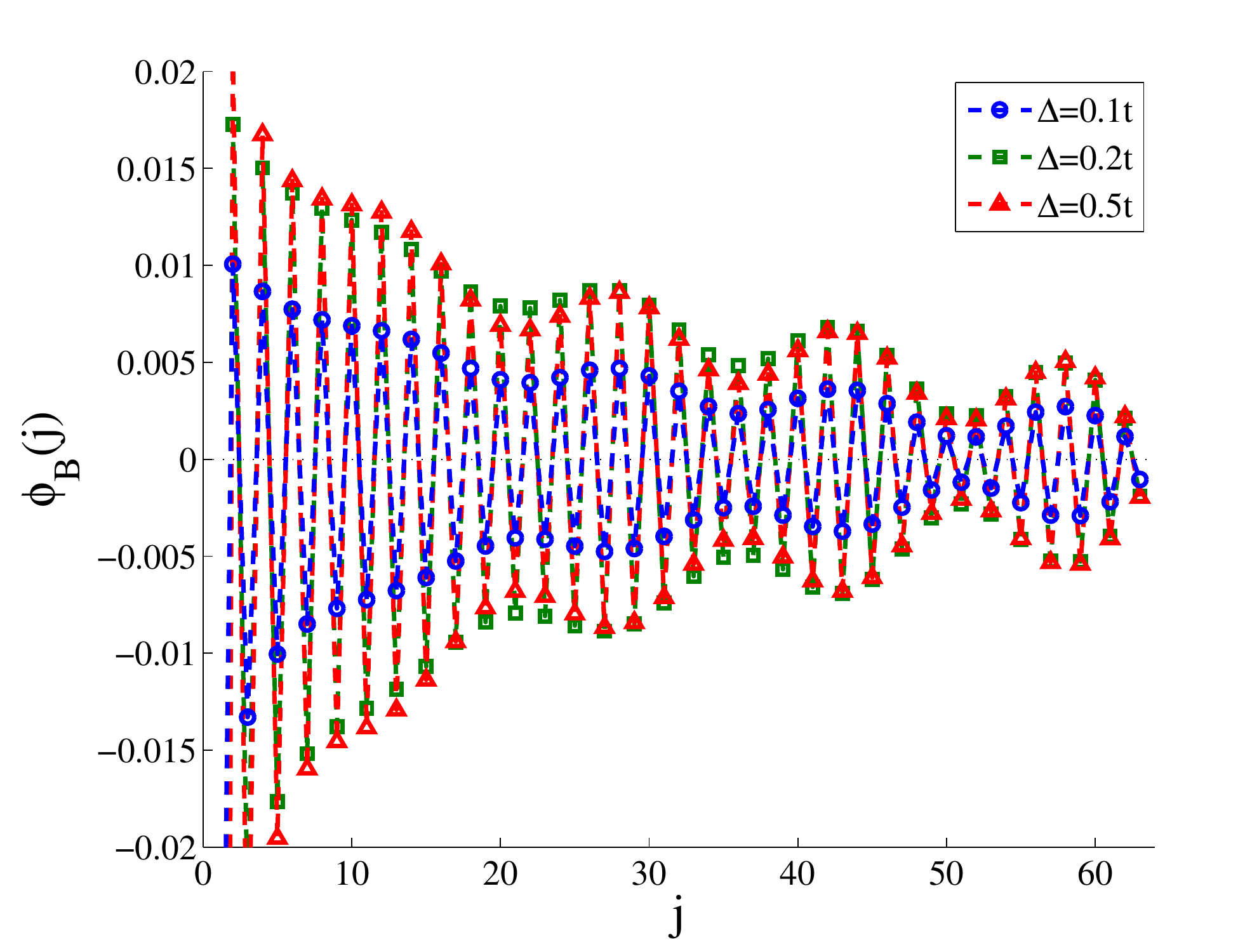}
\caption{Induced superconducting order parameters in $L=64$
systems with bond--centered boundary pair fields of varying
strength, as a function of position along the chain.} \label{fig1}
\end{figure}

\subsection{On--site boundary field}
In Fig. \ref{fig2}, we plot $\phi \left( j\right) $ and $\phi
_{B}\left( j\right) $ for an $L=64$ system in which the following
site--centered pair field was applied:
\begin{equation}
\tilde{H}_{\mathrm{pair}}=\Delta c_{1\uparrow }^{\dagger
}c_{1\downarrow }^{\dagger }+H.c.\text{,}
\end{equation}%
\begin{figure}[h]
\includegraphics[width=0.3\textwidth]{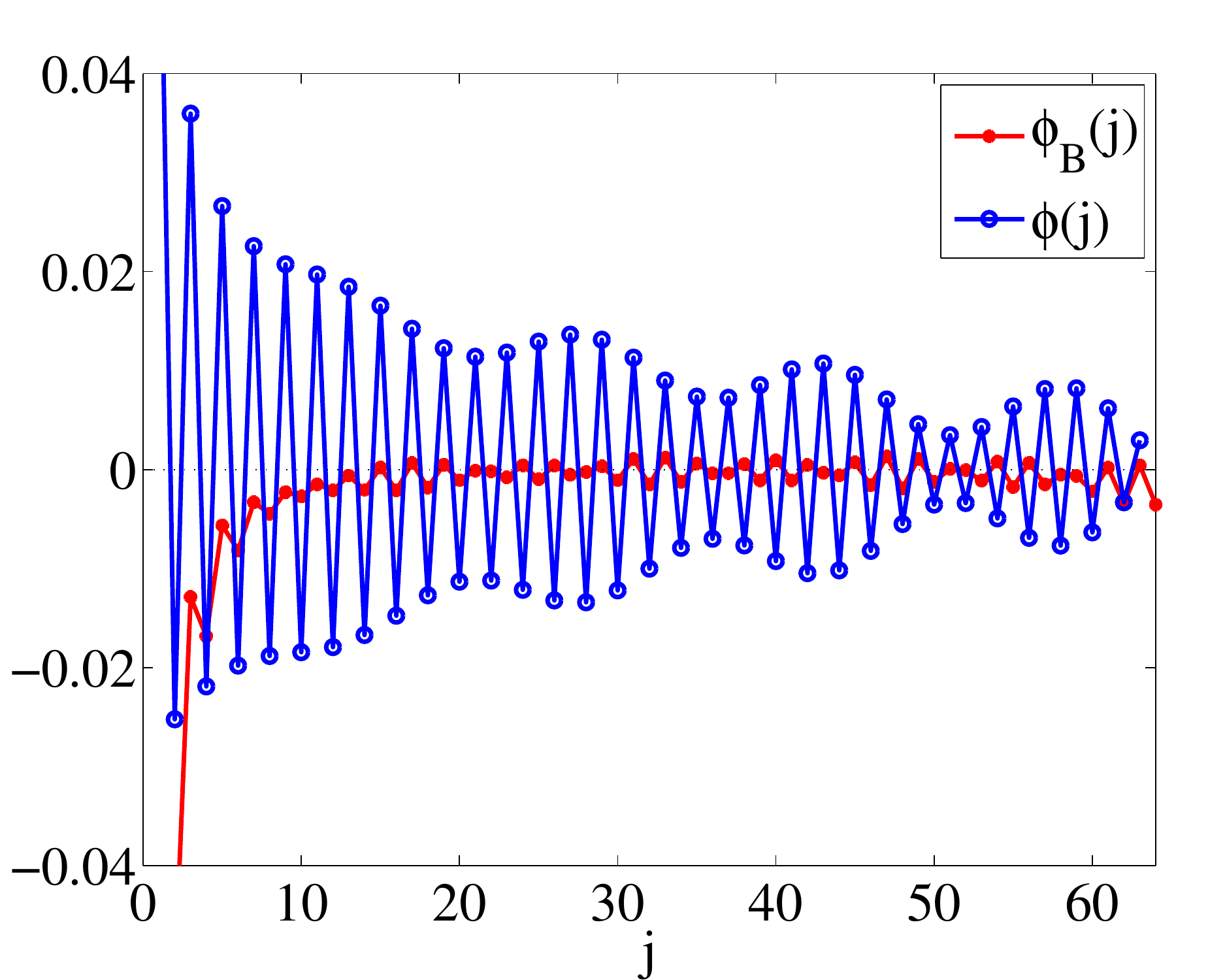}
\caption{Induced superconducting order parameters, $\phi$ and
$\phi_B$ (see Eq. \ref{OP}), in an $L=64$ system with a
site--centered boundary pair fields of strength $\Delta/t=0.5$.}
\label{fig2}
\end{figure}
with $\Delta =0.5t$. The results are very similar to the case of a
bond--centered boundary pair field: the on--site order parameter
$\phi \left( j\right) $ decays very rapidly, while a
bond--centered order parameter $\phi _{B}\left( j\right) $ is
induced. $\phi _{B}\left( j\right) $ oscillates at a wavevector
$q=\pi /a$ and its amplitude decays slowly. This shows that the
induced order parameter far away from the edges is not sensitive
to the details of the edge fields.

\subsection{Equal--time correlations}

Finally, we calculate the pair--pair correlation function
\begin{equation}
C\left( x\right) =\langle \phi _{B}\left( [L/2-x/2]\right) \phi
_{B}\left( [L/2+x/2]\right) \rangle \label{EQT}
\end{equation}%
in an $L=16$ system. $[\cdot]$ represents rounding to the nearest
integer from
below. The results (Fig \ref{fig3}) clearly show that beyond the first few neighbors, $%
C\left( x\right) $ oscillates with a period of $a$, i.e. the
dominant superconducting correlations are pair density wave
correlations, in agreement with measurements of the response to
edge fields.

\begin{figure}[h]
\includegraphics[width=0.35\textwidth]{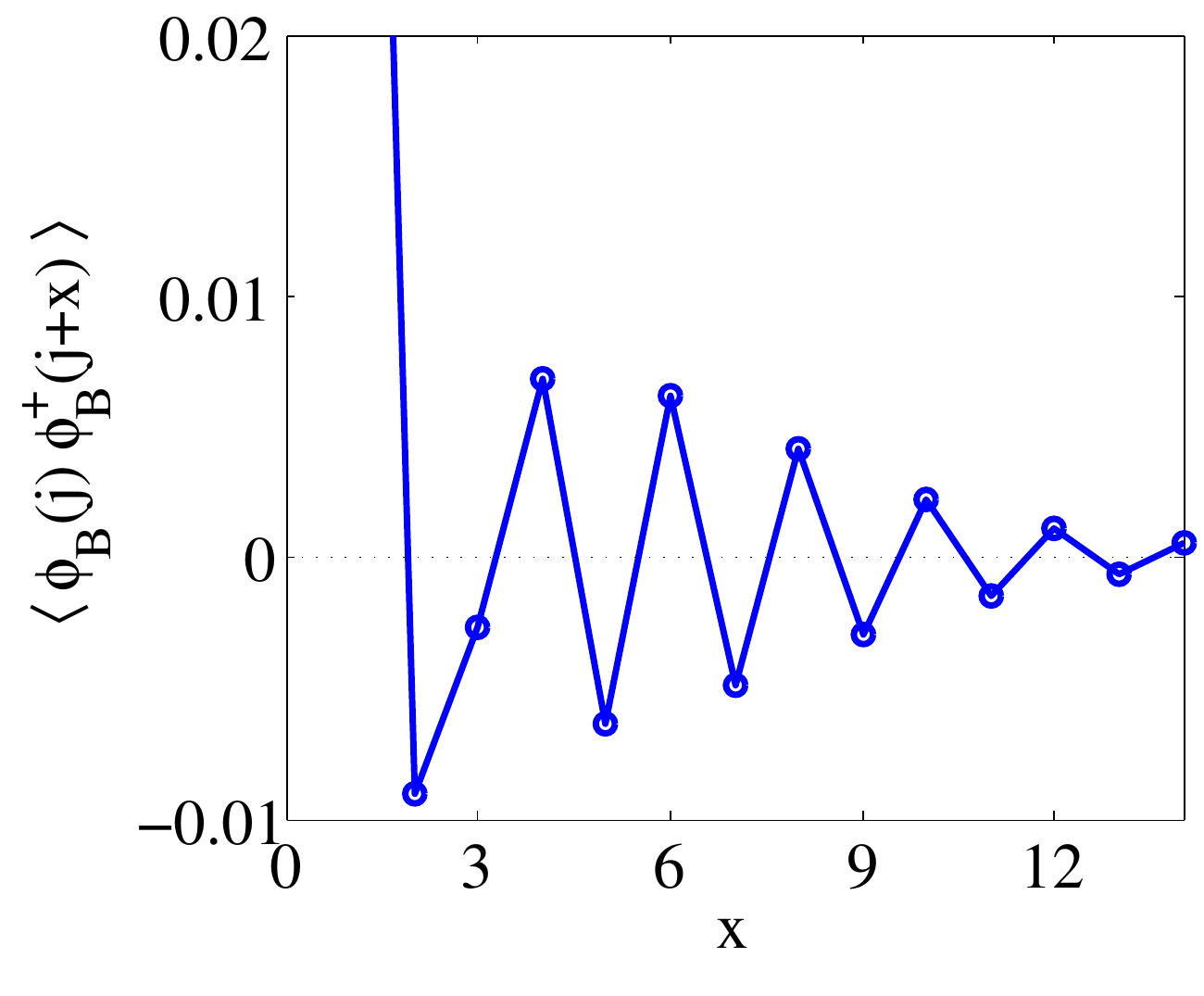}
\caption{The equal--time two--point superconducting correlation
function (Eq. \ref{EQT}) as a function of distance $x$.}
\label{fig3}
\end{figure}

\end{widetext}

\end{document}